\documentstyle[epsf]{aipproc}

\def\alwaysmath#1{\ifmmode{#1}\else{$#1$}\fi}

\def\teff{\alwaysmath{T_{\rm eff}}}

\begin{document}

\title{Constraining the Abundances and Ages of Early-type Galaxies}

\author{Ben Dorman$^{*\dagger}$ and Robert W. O'Connell$^*$}

\address{$^*$ Astronomy Department, University of Virginia\\ 
P.O. Box 3818, Charlottesville, VA 22903\\
$^{\dagger}$ UV/Optical Astronomy Branch, \\
Laboratory for  Astronomy \& Solar Physics, \\
Code 681, NASA/GSFC, Greenbelt, MD 20771}

\maketitle

\begin{abstract} We consider the extent to which broadband mid-UV
(2000 - 3000 \AA) colors can give superior constraints on the ages and
abundances of old stellar populations than can their optical
counterparts. The ultraviolet colors directly measure the turnoff
component of old populations.  They vary much more strongly with $t$
and $Z$ and give much tighter constraints for a given degree of
observational precision than do optical-band indices.  We present an
analysis of the permitted range of ($t,Z$) arising from observational
uncertainties in observations of early-type galaxies.

\end{abstract}

\section*{Introduction}

Am important problem in the study of stellar populations is that of
deriving ages and abundances for early-type galaxies. These are the
most massive galaxies in the universe and thought to be amongst the
oldest. In principle, therefore, their study offers one of the best
probes of the early star formation history and chemical evolution of
the universe.

\begin{figure}[ht]
\epsfxsize=4.5truein
\centerline{\epsffile{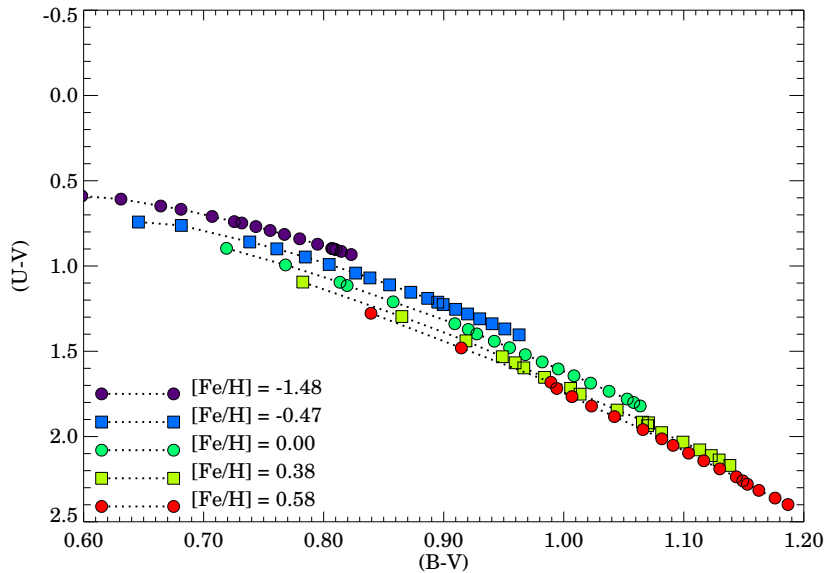}}
\vspace{10pt}

\caption{\protect\label{uvbv}
The two-color $(U-V),$ $(B-V)$ diagram as a function of age
and metallicity.  Model isochrones have been combined with Kurucz
\protect\cite{kur92} atmospheres as described in \protect\cite{dor97a}.  
The range of ages plotted is 2--20 Gyr with the 20 Gyr
models at the right-hand (red) end of the sequence and the models are
1 Gyr apart. [Fe/H] ranges from 1/30 to 3 $\times$ the solar value. }
\end{figure}

In this study, a progress report on our continuing exploration (see
\cite{dro93,dor95}) of old stellar populations at ultraviolet
wavelengths, we investigate the usefulness of mid-UV/optical broadband
colors.  A well-known problem in population synthesis is the
``age-metallicity degeneracy,'' (\cite{oc86,worthey94} and references
therein) by which populations of different age and abundance can have
nearly indistinguishable light measures ({\it viz.} colors, absorption line
strengths).  Broadband optical colors are particularly susceptible to this
problem.   Figure~(\ref{uvbv}) shows that observed integrated colors of,
for example, $(B-V) = 0.95,$ $(U-V) = 1.5$ are consistent with a wide
range of both ages and metallicities for typical observational uncertainties.
 
However, broadband colors have several important advantages over narrower
band measures, in particular

\begin{itemize}

\item Since they are influenced mainly by regions of low atmospheric
optical depth, they are easier to model theoretically than are
individual features --- which are often subject to poorly modeled effects;

\item It is easier to achieve high S/N ratios in broader indices;

\item They are not strongly affected by factors such as velocity
dispersion or emission features.

\end{itemize}

In the absence of young stellar components and the hot component that
gives rise to the `UV-upturn' or `UVX' phenomenon \cite{dor95,b3fl},
the mid-UV flux is dominated by the light of main sequence turnoff
stars. The turnoff light is much more sensitive to age than is the
light of red-giant stars, which dominate at optical and IR
wavelengths.  The UV is also more sensitive to stellar temperature and
line blanketing than are longer wavelengths.  Hence, we expect
indicators derived from the mid-UV to be considerably better probes of
population characteristics.  We show that this is true for mid-UV broad-band
colors, and we then address the question of how much photometric
precision is necessary to draw conclusions from these such measurements.

\section*{Mid-UV/Optical Colors}

\begin{figure}[ht]
\epsfxsize=4.5truein
\centerline{\epsffile{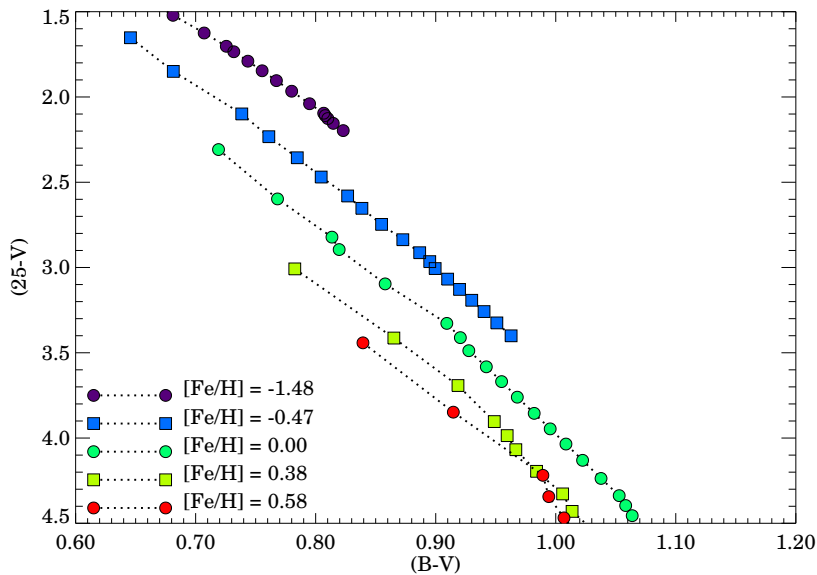}}

\vspace{10pt}
\caption{\protect\label{25vbv} Two-color diagram generated from the HST/WFPC2 color ${\rm
F255W}-V$ {\it vs.} $(B-V).$ The plot here is drawn to the same scale
as Fig.~\protect\ref{uvbv} and shows improved $t/Z$ separation.}
\end{figure}

We have used the HST/WFPC2 filters F218W, F255W, and F300W
 to exemplify the behavior of broadband mid-UV indices. These filters
have central wavelengths and effective widths defined in \cite{WFPC2}
as follows F218W -- $\lambda_{\rm peak} = 2091{\rm\, \AA},$ $\Delta
\lambda = 356 {\rm\, \AA};$ F255W -- $\lambda_{\rm peak} = 2483 {\rm\,
\AA},$ $\Delta \lambda = 408 {\rm\, \AA};$ F300W (the ``wide U'' used in
the Hubble Deep Field exposures)
$\lambda_{\rm peak} = 2760 {\rm\, \AA},$ 
$\Delta \lambda = 728 {\rm\, \AA}.$  The potential
utility of mid-UV broadband colors is illustrated by
Figure~\ref{25vbv}.  The two-color separation between model
sequences at different $Z$ is seen to be much greater in $(25-V)$ than
in $(U-V).$  In particular, the model pair (6 Gyr, 3$Z_{\odot}$) and (16 Gyr,
$Z_{\odot}$), cited by Worthey \cite{worthey94} as indistinguishable at
optical wavelengths, is well separated in the UV/optical colors (a full
difference spectrum is shown in \cite{dor97b}). Some ranges of
metallicity, or significant minority fractions thereof, are excluded
altogether by existing broadband observations of E galaxies.  For
example, for the galaxies presented in Table 1 of \cite{dor95}, $3.0 <
(25-V) < 3.7;$ this excludes most of the models with [Fe/H] $< -0.3$.
An even greater separation between models is present in the
$(22-V),(B-V)$ two-color plane.

However, note that the model sequences for constant age or constant
$Z$ still overlap in color---implying that the ``degeneracy'' problem
is still present, if less serious than in the optical.  In addition,
UV fluxes cannot yet be measured with the same precision as optical
fluxes due to remaining calibration problems and the faintness
of galaxies in the UV.

\section*{The Degeneracy Problem}

In this section we formulate a criterion for the distinguishability of
stellar population parameters from a given set of observations.  Let
$C$ be some measure of a galaxy spectral energy distribution (SED) such
as a color or absorption line index. Suppose the light-weighted mean
age and abundance of its population are $t_0$ and $Z_0.$ The range of
models that produce a value of $C$ within a given observational
uncertainty, $\sigma_{\rm obs}$,  is the set of $(t,Z)$ where

\begin{figure}[hp]
\epsfxsize=4.5truein
\centerline{\epsffile{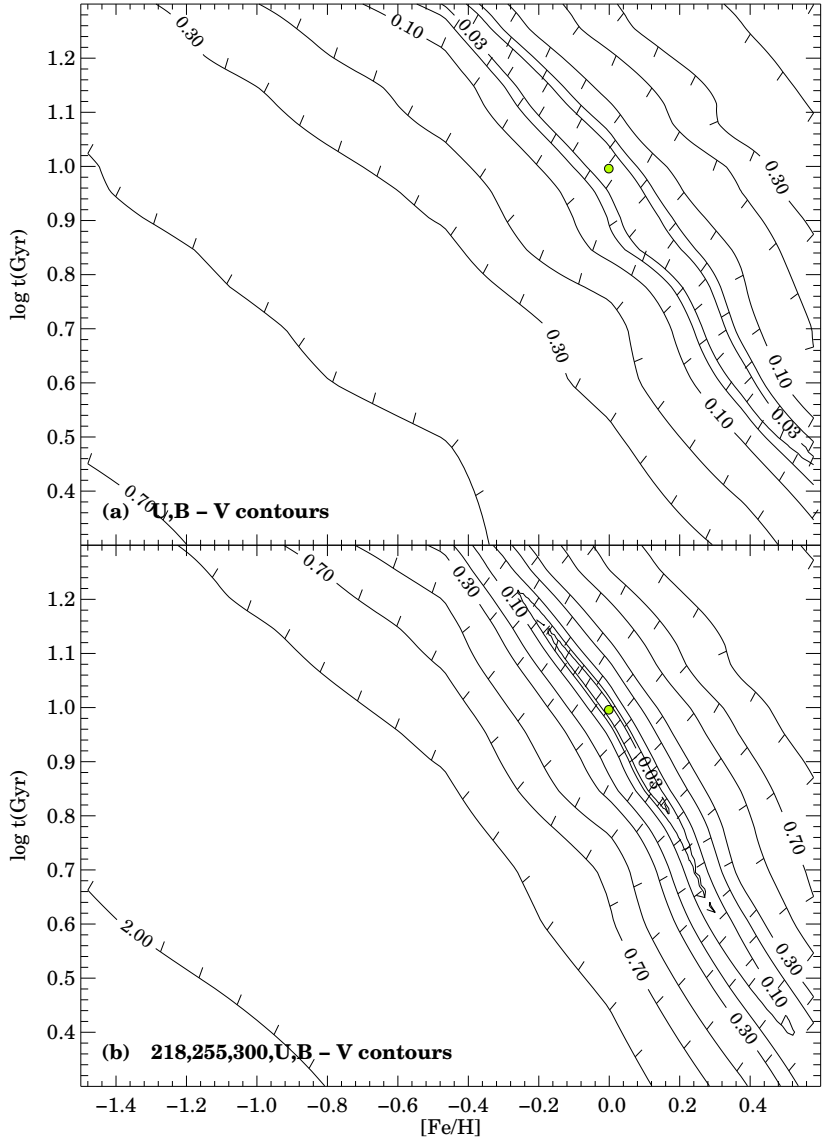}}

\vspace{10pt}
\caption{\protect\label{ellipse} RMS error ellipses generated from (a) the
optical $U,B,V$ bands,
and (b) these bands augmented by measurements in the HST F218W, F255W, F300W.
A filled circle denotes the 'input model,' and the contours give the envelope
of
models in the $(t,Z)$ plane that are indistinguishable from this input point
within
the contour value. Note the tightness of the contours in (b) compared to (a).}
\end{figure}

\begin{equation}
\label{sig}
\Delta C = \Big| C(t,Z) - C(t_0,Z_0) \Big| < \sigma_{\rm obs}.
\end{equation}

For a set of $n$ indices $\{C_i\},$ the range of models consistent
with the observation is the {\it error ellipse} in the $t,Z$ plane
defined by the r.m.s.  differences in colors between theory and
observation is determined by 

\begin{equation}
\label{r}
R =  \sqrt{ {{1}\over {n}} \sum_{i=1}^{n}\left[\Delta C_i\right]^2 } < 
\sigma_{\rm obs}.
\end{equation}

Using the partial derivatives of $C$ with respect to $t$ and $Z$ we
obtain an expression for $\Delta C$ in terms of the relative
sensitivity of the colors to these parameters:

\begin{equation}
\label{final}
\Delta C = \Bigg| {\Bigg({\partial \log C \over \partial \log t}\Bigg) }_Z d \log t  +
{\Bigg({\partial \log C \over \partial \log Z}\Bigg) }_t d \log Z \Bigg|,
\end{equation}

\noindent where the derivatives are both positive in the age range of
interest (colors become redder with both $t$ and $Z$\footnote{This
is not always true of absorption  indices, which may peak 
at a given $\rm \teff([Fe/H])$ owing to line saturation. In this case
equation~(\protect\ref{final}) must be modified to include higher-order derivative
terms.}). Worthey
\cite{worthey94} has defined ``metallicity sensitivity'' indices in
terms of the ratios (i.e.\ relative sizes) of the partial derivatives
in this expression.  However, it is important to note from equations 
(\ref{sig}) and (\ref{final}) that $t,Z$ separation also depends on the absolute
size of the partial derivatives and on the available observational
precision $\sigma_{\rm obs}$ (which determines the size of the error
ellipse).  

Figure~\ref{ellipse} shows the error ellipse contours of $R$ as defined
in equation~(\ref{r}). To construct this plot, we have selected a model
(10 Gyr, [Fe/H] = 0, indicated by the filled circle) and determined its
rms separation from other models in various colors\footnote{Note 
that this analysis addresses a different question than
the uncertainty analysis recently provided by
\cite{cwb96}, who addressed instead the systematic uncertainties inherent in 
the models themselves (i.e., how the contours in these diagrams would change
with different authors). Needless to say, both contribute to the uncertainty of
the results from population synthesis.}. In panel (a) the contours are derived
using only $(U-V), (B-V),$ while panel (b) is constructed using mid-UV/optical
colors in addition.

Even for the highest precision plotted, there is a ``valley of
degeneracy'' running diagonally across both figures wherein $t,Z$
separation is impossible.  This is the ``3/2'' band discussed by
Worthey \cite{worthey94}.  It is much broader in the optical than in
the UV bands.  The innermost contour plotted in both panels is 0.03
mag, a precision that can be routinely achieved for nearby galaxies in
the optical but not currently in the UV.  However, note that for
optical bands, as stressed by Worthey and his collaborators, the
degree of observational precision required to separate models is much
better than 0.03 mag.  The UV colors yield better separation for
$\sigma_{\rm obs} = 0.1$ than do the optical colors at $\sigma_{\rm
obs} = 0.03$.  

In practice, the mid-UV will often suffer measureable contamination
from the UVX, \cite{b3fl}.  However, this
is easily removed, as discussed in our forthcoming paper \cite{dor97a}.  

\section*{Discussion}

The age-abundance degeneracy problem is difficult and persists at some
level in all wavelength regimes which have been studied.  In the case
of the mid-UV colors we discussed here, degeneracy is much smaller at
a given level of observational precision than at optical wavelengths.
If there is independent information on abundance, constraints on age
from mid-UV colors are strong (see Fig.~\ref{ellipse}).  

It is clear from this discussion that the usefulness of various
spectral indices must be judged in terms of both their sensitivities
(i.e.\ partial derivatives) to the basic $t,Z$ parameters and the
likely observational precision possible in practical applications.  We
doubt that robust $t,Z$ separation is possible for old stellar
populations without excellent observational precision in multiple
spectral indices or colors over a long wavelength baseline (including
the mid-UV) $\cite{oc96}$.  It is also essential to evaluate the
systematic errors introduced by modeling uncertainties \cite{cwb96}.
We will discuss these issues further in our forthcoming paper
\cite{dor97a}.

This work has been partially supported by grants NASA grants
NAG5-700 and NAGW-4106.

\end{document}